# A test of optimal laser impulsion for controlling population within the $N_s$=1, $N_r$=5 polyad of $^{12}C_2H_2$

L. Santos[1], N. Iacobellis[1],
M. Herman[1], D.S. Perry[2],
M. Desouter-Lecomte[3,4], N. Vaeck[1]

[1]Laboratoire de Chimie quantique et Photophysique, CP160/09, Faculté des Sciences, Université Libre de Bruxelles; 50, ave. F. Roosevelt, B-1050, Belgium

[2] Department of Chemistry, The University of Akron, OH 44325-3601, USA

[3]Laboratoire de Chimie Physique, UMR8000, Univ Paris-Sud, University Paris-Saclay, Orsay, France

[4] Département de Chimie, Université de Liège, Bât B6c, Sart Tilman, B-4000 Liège, Belgium

contact person: N.Vaeck, nvaeck@ulb.ac.be

**Abstract**

Optimal control theory has been employed to populate separately two dark states of the acetylene polyad $N_s = 1$, $N_r = 5$ by indirect coupling via the ground state. Relevant level energies and transition dipole moments are extracted from the experimental literature. The optimal pulses are rather simple. The evolution of the populations is shown for the duration of the control process and also for the field free-evolution that follows the control. One of the dark states appears to be a potential target for realistic experimental investigation because the average population of the Rabi oscillation remains high and decoherence is expected to be weak.

## 1. Introduction

Since the development of the optimal control theory and its success in many experimental applications even in complex systems [1], there have been many theoretical attempts either to deduce control mechanisms [2] or to suggest new experimental investigations. However to be realistic, theoretical simulations require a very precise molecular Hamiltonian and efficient wave packet propagation methods so that such simulations usually concern atoms and small molecules [3-6] or control in reduced dimensionality [7, 8]. Control in a real environment still remains a challenging topic.

Up to now, this type of investigation was rarely applied to realistic vibrational case studies in full dimensionality. In the present work, we focus on acetylene ($^{12}C_2H_2$), taking advantage of the existence in the literature of a global Hamiltonian describing its vibrational and rotational energy levels within experimental precision, up to the near-infrared spectral range [9, 10]. The transition dipole moments, which are essential for the optimal control, have also been extracted from the experimental data. This global Hamiltonian is based on the use of polyads [11-13] gathering interacting zero-order vibrational wavefunctions into Hamiltonian matrices. The polyads are defined using the $N_s$ and $N_r$ pseudo quantum numbers, with $N_s = v_1+v_2+v_3$ and $N_r = 5v_1+3v_2+5v_3+v_4+v_5$. The quantum numbers $v_i$ provide the excitation level of vibrational normal modes with $i = 1$ (symmetric CH stretching), 2 (symmetric CC stretch), 3 (asymmetric CH stretch), 4 (twofold degenerate *trans*-bend) and 5 (twofold degenerate *cis*-bend). This polyad Hamiltonian has already been applied both to thermodynamics [14] and dynamics [15, 16]. The present test investigation deals with the polyad with $N_s = 1$ and $N_r = 5$, containing a minimal set of vibrational energy states to exhibit non-trivial dynamical behavior.

Our aim is to increase the population of a dark state, which would otherwise remain weakly populated, so that it could be probed experimentally, for example by time-resolved spectroscopy. Therefore, an optical pulse could be designed to maximize the population of that dark state starting from the groundstate (GS) with a very high performance index at the end of the pulse. However, as a dark state is a nonstationary state, it is a convenient target for optimal control only when its average population remains very high during the field free coherent evolution after the control pulse.

The paper is organized as follows. The polyad Hamiltonian matrix is detailed in section 2. Section 3 presents the target states for the optimal control in the eigenstate basis set. The optimal control method is explained in the section 4. Section 5 shows the result of this work and the conclusion is presented in section 6.

**2. Global acetylene Hamiltonian**

Extensive investigations of the $^{12}C_2H_2$ vibration-rotation spectrum under high-resolution spectroscopic conditions have led to the construction of a spectroscopic Hamiltonian. The version selected here [9] reproduces all 15,562 data published at the time within three times their experimental measurement precision. This precision was typically between $10^{-4}$ and $10^{-3}$ cm$^{-1}$. Vibrational states up to 8600 cm$^{-1}$ were considered in this investigation, based on a polyad perspective. For completeness, one should point out that more recent versions of the global Hamiltonian were since published [10,15], accounting for more data and interactions and over a broader energy range, however not affecting the present results on a low energy polyad. The literature cited in the introduction provides further information on this global Hamiltonian. Here, we shall consider only the polyad with $N_s = 1$ and $N_r = 5$ that contains the CH stretch fundamental vibration ($v_3 = 1$). This polyad (Eq.(1)) contains a minimal set of vibrational energy states that could exhibit non-trivial dynamical behavior.

$$\mathbf{H}_{N_s=1, N_r=5} = \begin{pmatrix} E_{(0010^00^0)} & a & b \\ a & E_{(0101^11^{-1})} & c \\ b & c & E_{(0101^11^1)} \end{pmatrix} \quad (1)$$

The three vibrational states in this polyad, $(v_1 v_2 v_3 v_4^{l_4} v_5^{l_5}) = (0010^00^0)$, $(0101^11^1)$ and $(0101^11^{-1})$, are all near 3300 cm$^{-1}$. The quantum numbers $l_4$ and $l_5$ are the vibrational angular momenta for the degenerate $v_4$ and $v_5$ normal modes of vibration. Their sum is $|k| = |l_4 + l_5|$. The adopted convention for the sign of $l_4$ is such that those states with $l_4 \geq 0$ and $l_4 < 0$ correspond to *e–* and *f–*symmetry states, respectively, as defined in the spectroscopic literature [17]. Interaction matrix elements connect only *e–* or *f–*states and only *g* or *u* states. The three vibrational states in Eq.(1) are the only states of *e* and *u* symmetries with $N_s = 1$ and $N_r = 5$. The zero-order states are connected by various resonances as detailed in Eqs.(1) to (4). Diagonal elements representing the zero-order energy include the conventional vibrational and rotational terms as well as, whenever relevant, the *l*-doubling terms as defined in the literature [12, 13, 18].

The interaction matrix element $a$ in Eq.(1) corresponds to a 3/245 anharmonic resonance, thus obeying the $\Delta v_3 = \pm 1$, $\Delta v_2 = \Delta v_4 = \Delta v_5 = \mp 1$ selection rule, with $\Delta k = 0$. It is defined, with $K_{3/245}$ the interaction constant, as:

$$\langle v_1, v_2, v_3, v_4^{l_4}, v_5^{l_5} | H | v_1, v_2-1, v_3+1, (v_4-1)^{l_4 \pm 1}, (v_5-1)^{l_5 \mp 1} \rangle = \tfrac{1}{8} K_{3/245} \sqrt{v_2(v_3+1)(v_4 \mp l_4)(v_5 \pm l_5)}$$

(2)

A second order interaction included in the global Hamiltonian has selection rules similar to 3/245 except that $\Delta k = \pm 2$. The $b$ interaction term in Eq.(1) is thus

$$\langle v_1, v_2, v_3, v_4^{l_4}, v_5^{l_5} | H | v_1, v_2-1, v_3+1, (v_4-1)^{l_4 \pm 1}, (v_5-1)^{l_5 \pm 1} \rangle = \tfrac{1}{8} O_{3/245} \sqrt{v_2(v_3+1)(v_4 \pm l_4)(v_5 \pm l_5)}$$
$$\times \sqrt{J(J+1) - k(k \pm 1)} \times \sqrt{J(J+1) - k(k \pm 1)(k \pm 2)}$$

(3)

The latter interaction ($O_{3/245}$ = 5.129 cm$^{-1}$) is much smaller than the former ($K_{3/245}$ = -18.300 cm$^{-1}$) and also $J$-dependent, where $J$ is the rotational quantum number and $J \geq |k|$. Actually, other $J$-dependent interactions of Coriolis-type were also included in the final global Hamiltonian [10], however they are not relevant to the present case study. Those interaction constants are effective parameters determined in Ref. [9]

The matrix element $c$ in Eq.(1) represents the rotational $l$-doubling interaction, with the subscript $b$ referring to one of the degenerate bending modes $\nu_4$ and $\nu_5$:

$$\langle l_b | H | l_b \pm 2 \rangle = \tfrac{1}{4} q_b \sqrt{(v_b \mp l_b)(v_b \pm l_b + 2)} \\ \times \sqrt{J(J+1) - k(k \pm 1)} \times \sqrt{J(J+1) - k(k \pm 1)(k \pm 2)} \quad . \quad (4)$$

The latter two interaction terms, $b$ and $c$, are strongly $J$-dependent, increasing roughly as $J^2$. Therefore a different Hamiltonian matrix applies for each value of $J$. Also the interaction terms are parity-dependent in $k$. Only even $k$ values (= 0 and 2) are involved in the Hamiltonian matrix considered in Eq.(1).

In order to appreciate the relative size of the various matrix elements, the effective Hamiltonian matrix is detailed in cm$^{-1}$ in Eq.(5) for the present case study, i.e. $J = 30$.

$$\mathbf{H}_{N_s=1, N_r=5, J=30} = \begin{pmatrix} 4375.800918 & 6.393881 & 0.039923 \\ 6.393881 & 4378.868219 & -4.669548 \\ 0.039923 & -4.669548 & 4392.943614 \end{pmatrix} \quad (5)$$

In this preliminary report, we have not addressed rotation beyond selecting an upper rotational level with $J$ value high enough to have the corresponding polyad displaying significant off-diagonal interaction matrix elements. Actually, addressing feasible experimental transitions, i.e. those fitting $\Delta J = \pm 1$ selection rules, would require accounting for the GS starting rotational level. Also some rotational dependence in the transition dipole should be accounted for in future developments. Here, the GS starting rotational level has been imposed to be $J = 0$, not affecting the present results by any means.

## 3. Eigenstates and target states

In order to search for an optimal pulse for moving population into the dark states of the $J = 30$ polyad from the GS, one must first determine the eigenenergies and the eigenstates of this subspace that is decoupled from GS. The Hamiltonian matrix (Eq.(5)) is diagonalized to get the eigenenergies: $\Gamma_1 = 4370.36$ cm$^{-1}$, $\Gamma_2 = 4382.7$ cm$^{-1}$ and $\Gamma_3 = 4394.55$ cm$^{-1}$. The linear combinations of the zero-order states for the different eigenstates $|\Gamma_{1-3}\rangle$ are the following:

$$\begin{aligned}
|\Gamma_1\rangle &= 0.755351|0,0,1,0^0,0^0\rangle - 0.641475|0,1,0,1^1,1^{-1}\rangle - 0.133994|0,1,0,1^1,1^1\rangle \\
|\Gamma_2\rangle &= 0.646271|0,0,1,0^0,0^0\rangle + 0.69532|0,1,0,1^1,1^{-1}\rangle + 0.314427|0,1,0,1^1,1^1\rangle \quad . \quad (6)\\
|\Gamma_3\rangle &= -0.108528|0,0,1,0^0,0^0\rangle - 0.324099|0,1,0,1^1,1^{-1}\rangle + 0.939777|0,1,0,1^1,1^1\rangle
\end{aligned}$$

In the zero-order basis set, the only transition that can occur from the GS is the transition to the vibrational state $|0,0,1,0^0,0^0\rangle$, with a transition dipole moment $\mu_t$. This state is called the bright state and the other zero-order states, for which transitions from the GS are forbidden, are called dark states [11]. Note for the discussion to come that the transitions among zero-order states within the polyad are not allowed because of the electric dipole selection rule (u ← u forbidden). The value of $\mu_t$ is estimated from the vibrational contribution given in Ref. [19] that has been experimentally determined by high-resolution Fourier transform spectroscopy to be 0.08907 Debye. In the eigenstate basis set, the transition dipole moment matrix $D$ is the following:

$$\mathbf{D} = \begin{pmatrix} 0 & 0.755351\mu_t & 0.646271\mu_t & -0.108528\mu_t \\ 0.755351\mu_t & 0 & 0 & 0 \\ 0.646271\mu_t & 0 & 0 & 0 \\ -0.108528\mu_t & 0 & 0 & 0 \end{pmatrix}. \quad (7)$$

Now, the bright character of the bright state $|0,0,1,0^0,0^0\rangle$ is distributed among the three eigenstates, so now more than one transition is possible. However, the transitions between eigenstates are still not allowed.

In order to test the control of the population within the $N_s = 1$, $N_r = 5$ polyad, we chose to populate the two dark states of the polyad, $|0,1,0,1^1,1^1\rangle$ and $|0,1,0,1^1,1^{-1}\rangle$, after a time *T*. These states are the different target states for the optimal control process. In the eigenstate basis set, the two target states denoted $|\phi_+\rangle$ and $|\phi_-\rangle$ correspond to the following linear combination of the eigenstates, respectively:

$$|\phi_+\rangle = |0,1,0,1^1,1^1\rangle = -0.133994|\Gamma_1\rangle + 0.314427|\Gamma_2\rangle + 0.939777|\Gamma_3\rangle$$
$$|\phi_-\rangle = |0,1,0,1^1,1^{-1}\rangle = -0.641475|\Gamma_1\rangle + 0.69532|\Gamma_2\rangle - 0.324099|\Gamma_3\rangle \quad . \tag{8}$$

Since the control begins with the whole GS population, i.e., the initial state of the control, $|\psi(0)\rangle$, there are two optimal processes. Each of them corresponds to the total transfer of the population to one of the zero-order dark states of the polyad. These targets are obviously non-stationary states and will give rise to an oscillatory behavior. This oscillatory behavior could be destroyed by the coupling with external states such as states from other polyads that act as dissipative bath. Hovewer, in the polyad model, the coupling between polyads is by definition very weak at this range of energy [20, 10, 21] and was not included in the simulation. So if the Rabi oscillation amplitude is weak, the population in the dark state can remain sufficiently high to allow for an experimental probe.

**4. Optimal field design**

Optimal control theory [22] is used to search for the optimal laser pulse that corresponds to this transfer of population. We consider a linearly polarized field so the Hamiltonian for the control is given by $H(t) = H_0 - D\varepsilon(t)$ where $H_0$ is the diagonal matrix containing the GS and the three eigenstates of the polyad and $D$ is the transition dipole moment matrix in the direction of the field $\varepsilon(t)$ (see Eq.7). This field is obtained by maximizing a functional $F$ built on an objective and constraints. Here, the objective $I$ is the probability of reaching the target at the final time when the system is driven by the optimal field. $I$ is thus the square modulus of the overlap between the final state superposition of the optimal process $|\psi(T)\rangle$ and the target state superposition $|\phi\rangle$,

$$I = \left|\langle\psi(T)|\phi\rangle\right|^2. \tag{9}$$

Even so, the transition between the initial state $|\psi(0)\rangle$ and the final state $|\psi(T)\rangle$ is known up to a phase factor [22]. The objective is furthermore defined as the performance index of the optimal process and gives the accuracy of the optimal field. Here, we take into account two constraints ($C_1$, $C_2$), one limits the total integrated intensity of the electric field,

$$C_1 = -\int_0^T \alpha(t)|\varepsilon(t)|^2 dt \ . \tag{10}$$

Here $\alpha(t) = \alpha_0 / \sin^2(\pi t / T)$ where $\alpha_0$ is a parameter that controls the strength of the constraint and the sine function forces the field to switch smoothly on and off [23]. The other constraint ensures that the Schrödinger's equation is always satisfied during the entire process [24],

$$C_2 = -2Re\left\{\langle\psi(T)|\phi\rangle \int_0^T \langle\lambda(t)|\frac{i}{\hbar}H + \frac{\partial}{\partial t}|\psi(t)\rangle dt\right\} \qquad (11)$$

where $\lambda(t)$ is the Lagrange multiplier for the Schrödinger's equation constraint. The functional $F$ is the sum of the three previous terms,

$$F = |\langle\psi(T)|\phi\rangle|^2 - \int_0^T \alpha(t)|\varepsilon(t)|^2 dt - 2Re\left\{\langle\psi(T)|\phi\rangle \int_0^T \langle\lambda(t)|\frac{i}{\hbar}H + \frac{\partial}{\partial t}|\psi(t)\rangle dt\right\}. \qquad (12)$$

Note that other experimental constraints could be added, for example a limited spectral range [25] or a zero integrated area of the pulse [5]. This functional $F(\psi, \lambda, \varepsilon)$ is maximized with respect to variation of $\psi, \lambda, \varepsilon$. This leads to three equations. The two first equations are time-dependent Schrödinger equations. The first applies to the initial condition $\psi(t=0) = \psi(0)$ and the second to the final condition with the Lagrange multiplier $\lambda(T) = \phi$. Finally, the field $\varepsilon(t)$ is obtained from the corresponding wave packets at any time $t$. The expression for the field is

$$\varepsilon^*(t) = -\frac{1}{\hbar\alpha(t)} Im\left\{\langle\psi(t)|\lambda(t)\rangle\langle\lambda(t)|\mu|\psi(t)\rangle\right\}. \tag{13}$$

The three coupled evolution equations are solved by the Rabitz monotonous convergent iterative algorithm [24] with small time steps *dt*. After each iteration *i*, the field is obtained by $\varepsilon_i(t) = \varepsilon_{i-1}(t) + \varepsilon_i^*(t)$ where $\varepsilon_i^*(t)$ is given by Eq.(13) for each iteration. Note that the algorithm begins with an initial guess field. Here, we choose a very simple field with sine square envelopes,

$$\varepsilon_0(t) = \sum_i^N A \sin^2\left(\frac{\pi t}{\sigma}\right)\cos(v_i t) \tag{14}$$

where $A$ and $\sigma$ are respectively the amplitude and the width parameter of each sub-field and $v_i$ are the transition frequencies associated to the transitions between the GS and the eigenstates of the polyad. Since each sub-field has the same $\sigma$ duration, the duration of the field is equal to $\sigma$. Note that in the present case, $\sigma$ also equals the duration $T$ of the optimal process.

## 5. Results

The guess field includes three sub-fields, one for each transition from the GS to one of the eigenstates. Following Eq.(14) and including the duration of the optimal process $T$, the field is now given by

$$\varepsilon_0(t) = A \sin^2\left(\frac{\pi t}{T}\right)\left[\cos(v_1 t) + \cos(v_2 t) + \cos(v_3 t)\right] \quad (18)$$

where the transition wavenumbers are $\tilde{v}_1 = 4370.36$ cm$^{-1}$, $\tilde{v}_2 = 4382.7$ cm$^{-1}$ and $\tilde{v}_3 = 4394.55$ cm$^{-1}$. These transition wavenumbers correspond to the eigenenergies of the Hamiltonian matrix in Eq.(5).

Three parameters were varied in order to rapidly maximize the performance index $I$. The first parameter is the duration of the fields, $T$. In practice, we have chosen $T$ and then a time step in order to have a good temporal sampling. $T$ needs to be above the Heisenberg limit required in order to resolve the three states of the polyads. This limit given by $\Delta t \geq \hbar/2\Delta\Gamma$ (where $\Delta\Gamma$ is the energy gap between the eigenstates in the target polyad) is a consequence of the uncertainty principle and is around 0.2 picoseconds (ps) in the present case. The parameter $T$ has a direct influence on the performance index $I$. If the field is longer, the performance index will increase faster. A longer field looks then like a better choice but, since we are working with only one polyad, one must typically stay under the nanosecond scale in order to be able to neglect inter-polyad couplings [11]. The total duration of the fields $T$ is then fixed at 24.2 ps. This value for *T* is the same as the value used by Gruebele and coworkers in a similar energy range [26]. The integration time step $dt$ is 48.4 attoseconds. This time step is chosen for correct sampling of the period of the carrier frequencies in the guess field.

The second parameter is the penalty factor, $\alpha_0$. The higher this parameter is, the lower the amplitude of the optimal field will be and the slower the performance index will increase. With $T$ = 24.2 ps, $A$ = $10^6$ V/m and $\alpha_0$ = 1, the performance index increases rapidly and the field shape varies rapidly at each iteration. The amplitude of the optimal field for $|\phi_-\rangle$ is $2.18 \times 10^9$ V/m, high enough to compete with the intramolecular Coulomb forces. With $\alpha_0$ = 50, the amplitude of the optimal field is $8.40 \times 10^8$ V/m, which seems more acceptable. This value remains in the perturbative regime [27]. Indeed, Bandrauk *et al.* proposed to use the frequency of radiative transition, $\omega_R = (d \times E)/\hbar$, to determine the limit of the perturbative regime. In our case, $\omega_R \approx 10^8 s^{-1}$, is much smaller than the natural vibrational frequencies of our molecule.

The last parameter is the amplitude of the guess fields, *A*. The higher this parameter is, the faster the performance index *I* will increase. But it has a moderate influence on the performance index because it only controls the shape of the guess field. Moreover, a high amplitude *A* for the guess fields induces a high amplitude for the optimal fields. Consequently, the chosen value is $A = 10^7$ V/m in order to improve the performance index while maintaining a reasonable amplitude for the optimal fields.

The fields have been optimized up to a performance index $I$ above 0.99999 and such a high convergence requires about 200 iterations. The propagation of the dynamical equations is carried out by the fourth order Runge-Kutta method [28, 29] in the interaction representation.

The optimal fields $\varepsilon_+$ and $\varepsilon_-$ corresponding respectively to the target states $|\phi_+\rangle$ and $|\phi_-\rangle$ are shown in Figure 1. The maximal amplitude of the two optimal fields remains close to the amplitude of their guess fields: $8.40 \times 10^8$ V/m for $|\phi_+\rangle$ and $4.11 \times 10^8$ V/m for $|\phi_-\rangle$. The performance index $I$ is 0.99999 for $|\phi_+\rangle$ and 0.99999 for $|\phi_-\rangle$. It is the difference of coupling between the two target states and the bright state that induces the difference of $I$. Because of the nature of this coupling, the bright state population, which is prepared by the dipole-allowed transition from GS, goes more easily to the dark state $|\phi_-\rangle$ than to $|\phi_+\rangle$.

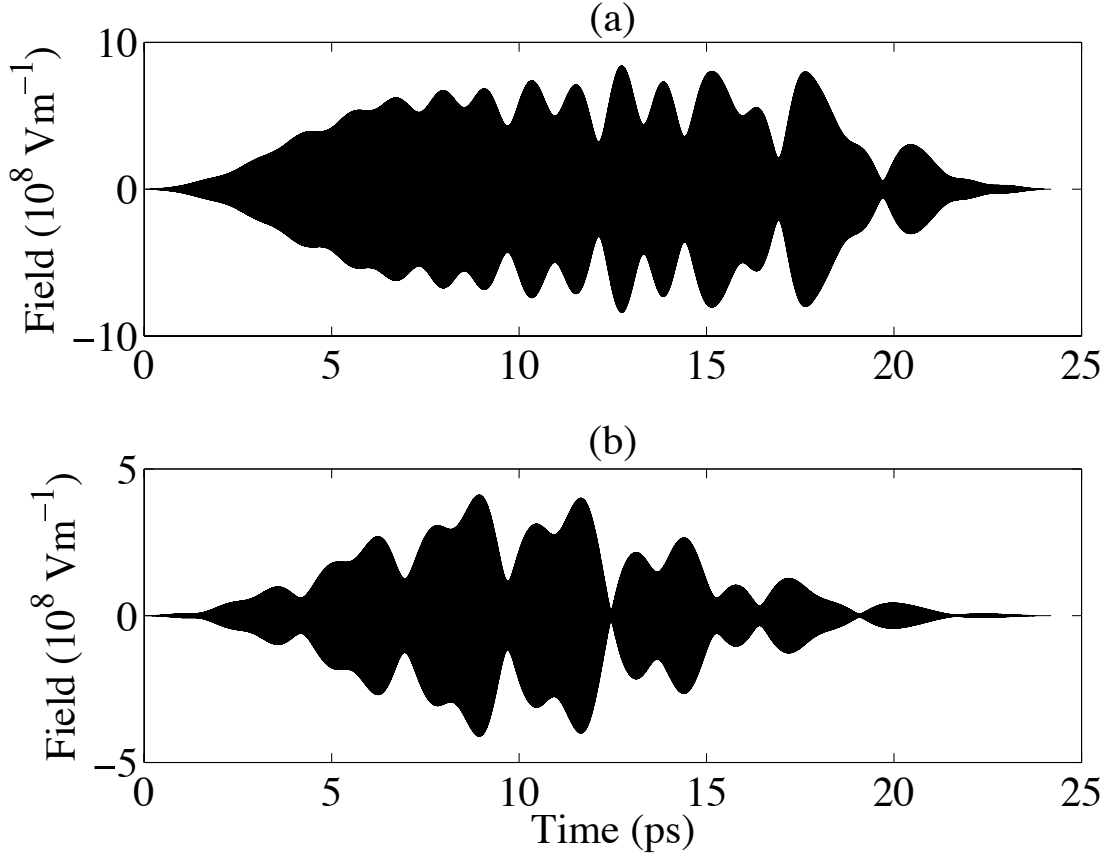

**Figure 1**: (a) The optimal field $\varepsilon_+$ for preparation of the target state $|\phi_+\rangle$ with a maximal intensity of $8.40 \times 10^8$ V/m and a performance index $I$ of 0.99999. (b) The optimal field $\varepsilon_-$ for the target state $|\phi_-\rangle$ with a maximal intensity of $4.11 \times 10^8$ V/m and a performance index $I$ of 0.99999. The duration $T$ of the fields is 24.2 ps.

The squared modulus of the Fourier transform $|S(\nu)|^2$ of the fields are shown in Figure 2. The spectra are similar in each case, with three peaks corresponding to the three guess field frequencies ($\nu_1, \nu_2, \nu_3$), but the relative intensities and contours of the peaks are different. For the field $\varepsilon_+$

, peak $v_3$ is very much larger than peaks $v_1$ and $v_2$, while for the field $\varepsilon_-$, peak $v_3$ is still the largest but the ratio is less dramatic. This can be explained by large overlap of the target state $|\phi_+\rangle$ to the mostly dark eigenstate $\Gamma_3$ (see Eq.(8)). Consequently, this state is less easily reachable and the control pulse needs more intensity to populate it. For the field associated to this target state (Fig. 2 (a)), there are also two small peaks at 4391 and 4397 cm$^{-1}$ just before and after the third peak. These peaks do not correspond to any transitions in the basis states. We have tried to remove these contributions from the optimal pulse but when we re-optimized the new pulse, the two peaks reappeared. So the initial optimal pulse, which includes these two small shoulder peaks, was kept. These side peaks correspond to the frequency shift from 4394.55 cm$^{-1}$ to 4391 and 4397 cm$^{-1}$ that occurs at the end of the control pulse (Figure 3). Gabor Transforms (GT) [30] were computed for the two fields, and the one for the $\varepsilon_+$ reveals interesting temporal features (figure 3). To obtain the graph, the pulse has been split in three time windows and the chosen time window is an exact Blackman window [31]. As shown in Figure 3, the peak for $v_3$ is present from the beginning to the end of the control pulse, but the peaks for $v_2$ and $v_1$ are respectively only present in the middle and at the end of the pulse. As in Figure 2, the peak for $v_3$ is more intense than the others.

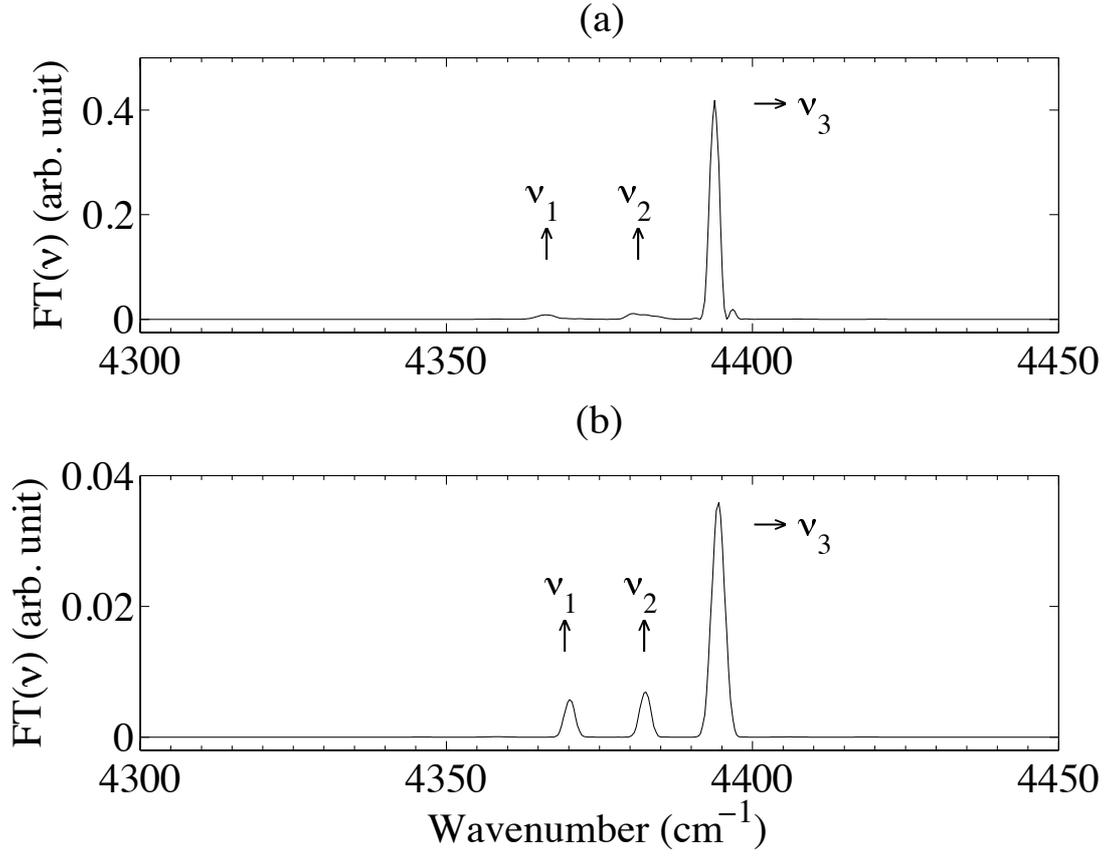

**Figure 2**: The squared modulus of the Fourier transform $|S(\nu)|^2$ of the field $\varepsilon_+$ (panel a) and of the field $\varepsilon_-$ (panel b). In both cases, the three peaks of the guess field frequencies are noticeable (in wavenumbers $\tilde{\nu}_1$ = 4370.36 cm$^{-1}$, $\tilde{\nu}_2$ = 4382.7 cm$^{-1}$ and $\tilde{\nu}_3$ = 4394.55 cm$^{-1}$).

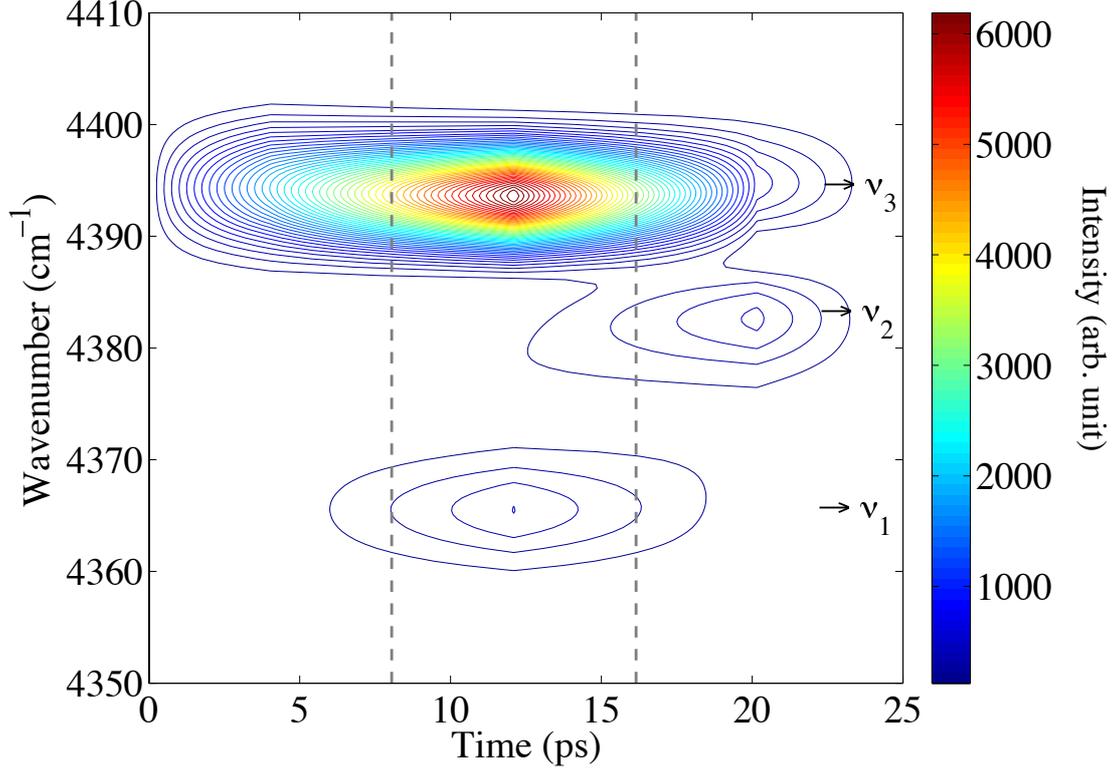

**Figure 3**: The Gabor transform of the field $\varepsilon_+$ obtained with three exact Blackman time windows. The peak corresponding to $\nu_3$ is always present and intense while the peaks for $\nu_1$ and $\nu_2$ are respectively present in the middle and at the end of the duration of the pulse.

It is instructive to look at the time-evolution of the populations of the eigenstates during the control pulses. The target populations of the eigenstates calculated from Eq.8 are shown in Table 1.

**Table 1**: Target populations of the optimal process for the target states $|\phi_+\rangle$ and $|\phi_-\rangle$.

|            | $\|\Gamma_1\rangle$ | $\|\Gamma_2\rangle$ | $\|\Gamma_3\rangle$ |
|------------|-----------|-----------|----------|
| $\|\phi_+\rangle$ | 0.0179544 | 0.0988643 | 0.883181 |
| $\|\phi_-\rangle$ | 0.411490  | 0.483470  | 0.105040 |

As shown in Figure 4, the respective target population is reached with either optimal field by the end of the control pulse. For the target state $|\phi_+\rangle$, the population of the eigenstate $|\Gamma_3\rangle$ increases monotonously while the population of the GS decreases with oscillations, and populations of $|\Gamma_1\rangle$ and $|\Gamma_2\rangle$ increase slightly. As explained in section 3, the transitions between eigenstates are not allowed. Consequently, the oscillations are attributable to the up and down transitions between the GS and the eigenstates $|\Gamma_1\rangle$ and $|\Gamma_2\rangle$ leading ultimately to the desired final eigenstate populations. The oscillations are stronger for the target state $|\phi_+\rangle$, which has a large mixing coefficient of the mostly dark $|\Gamma_3\rangle$ (see Eq.(8)).

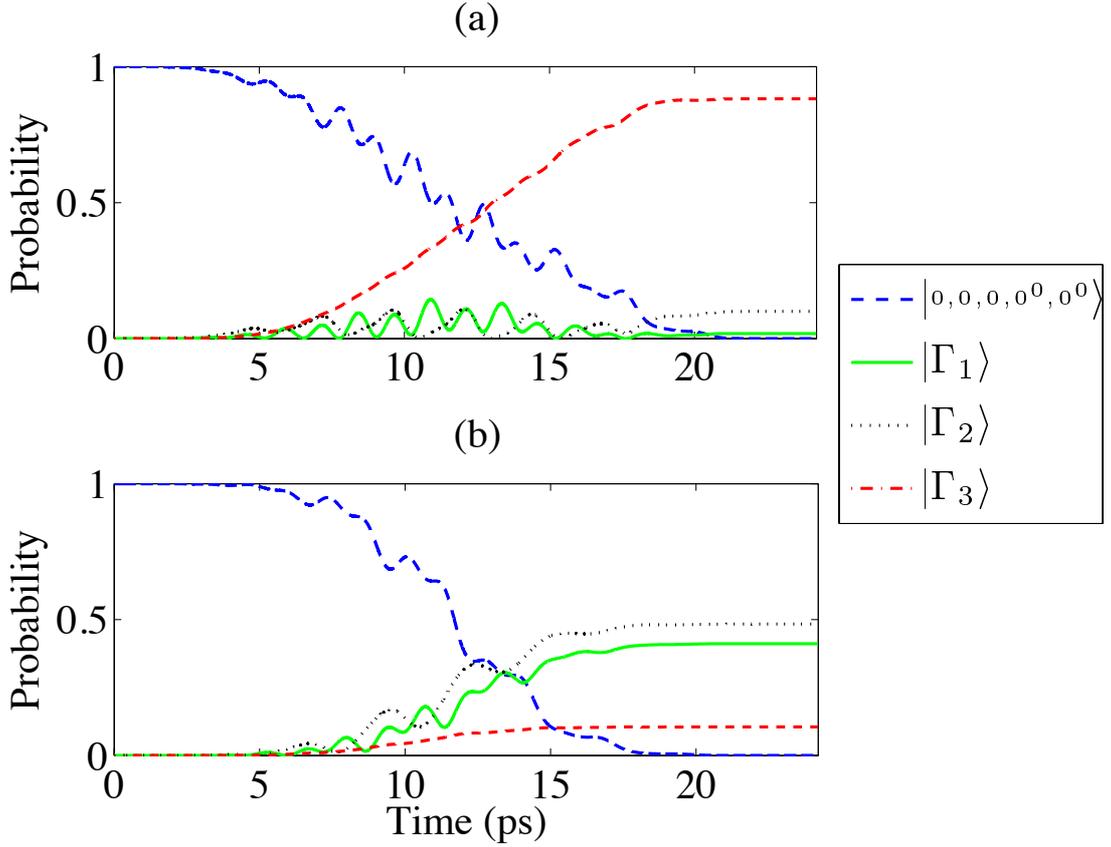

**Figure 4**: Evolution of the populations of the eigenstates for the field $\varepsilon_+$ (a) and for the field $\varepsilon_-$ (b).

By the application of optimal control, we have populated the zero-order states $|0,1,0,1^1,1^1\rangle$ and $|0,1,0,1^1,1^{-1}\rangle$, which are non-stationary states. The field-free evolutions of these states are shown in Figure 5. For both prepared zero-order states, there are oscillations between the three zero-order states of the polyad during the 121 ps of field-free time. These include a deeply modulated fast oscillation with a period near 3 ps and a weaker slow oscillation with a period of 67.6 ps. The fast oscillation reflects the strong coupling between the zero-order states, $|0,1,0,1^1,1^{-1}\rangle$ and

$|0,0,1,0^0,0^0\rangle$. The slow oscillation reflects the weak coupling of this strongly coupled pair with $|0,1,0,1^1,1^1\rangle$. The population of the target state $|0,1,0,1^1,1^1\rangle$ (panel a), remains high throughout the field-free evolution, so this state is proposed as an attractive target for practical optimal control experiments. On the other hand, the target state $|0,1,0,1^1,1^{-1}\rangle$ (panel b) shows rapid deep oscillations with $|0,0,1,0^0,0^0\rangle$ and the probability of $|0,1,0,1^1,1^{-1}\rangle$ decreases to nearly zero every 3 ps.

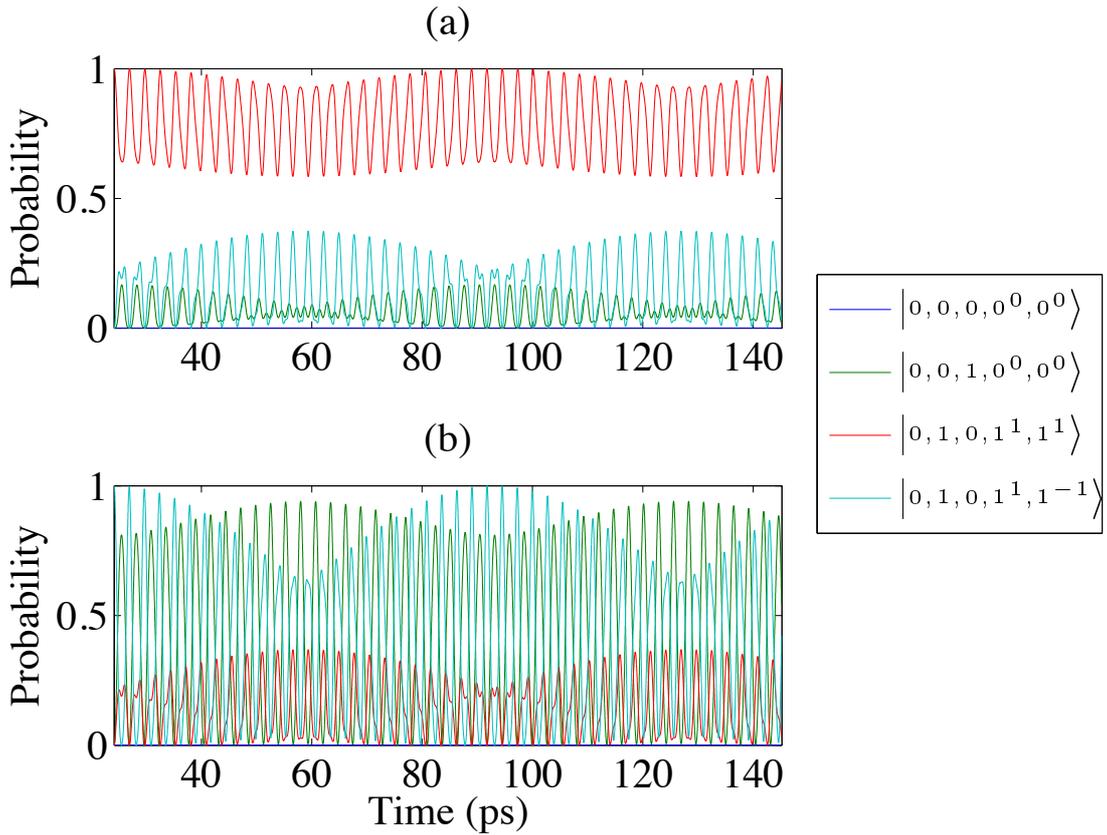

**Figure 5**: The dynamical free evolution of the zero-order state $|0,1,0,1^1,1^1\rangle$ (a) and $|0,1,0,1^1,1^{-1}\rangle$ (b). In both cases, the duration of the

dynamical free evolution is $T_{free}$ = 121 ps, beginning immediately following the control pulse at $T$ = 24.2 ps and ending at 145.2 ps on the horizontal scale. Note that the GS curve is not distinguishable from the x-axis on the graphs.

## 6. Conclusions

An optimal control calculation has been employed to populate two dark states of the acetylene polyad $N_s$ = 1, $N_r$ = 5. The dark state $|0,1,0,1^1,1^1\rangle$ is a potential target for experimental investigations because the average population is predicted to remain high during an extended period of field-free evolution after the control pulse. This assumes that the coupling to other polyads is weak enough to maintain the coherence created by the optimal control pulse. The derived quantum control pulses are rather simple suggesting that experimental implementation would be feasible. This work opens the way to higher polyads where pulses could populate pure bending ($N_s$ = 0) dark states. Such states are believed to be involved in the isomerization of acetylene to vinylidene [32].

To build on the present work, the next step would be to use the complete vibrational basis set for the polyad, including states coupled by Coriolis and rotational $l$-resonances, which will impact the efficiency of the optimal control field. Polyads that have similar total energies and also states accessed by hot band transitions [33] may also be considered.

Since these polyads are well isolated from other polyads, they could be employed for quantum computation. Following the pioneering work of Tesch et al., optimal pulses can be designed to serve as logical gates [34], and quantum beats may be used for the readout process [35].

**Acknowledgments**

This work was sponsored by the "Wiener-Anspach" foundation (ACME project) and supported by the FRS-FNRS of Belgium under Grant n° I.I.S.N. 4.4504.10. N.I. and L.S. acknowledge a F.R.I.A. research grant from the FRS-FNRS of Belgium. M.D-L. thanks the FRS-FNRS of Belgium for a grant. We also thank the COST XLIC and MOLIM actions. This work has been performed within the French GDR 3575 THEMS. DSP acknowledges support from the U.S. Department of Energy, Office of Basic Energy Sciences under Grant No. DE-FG02-90ER14151.